    TEX -- dialect:          revtex   3.0

\documentstyle[prb,aps]{revtex}
\tighten

\begin{document}
\setcounter{equation}{0}  \draft
%\twocolumn[                                    % full width abstract
 \twocolumn[\hsize\textwidth\columnwidth\hsize  % half width abstract
            \csname @twocolumnfalse\endcsname   % half width abstract

\title{UNIVERSALITY OF FREQUENCY AND FIELD SCALING OF THE
CONDUCTIVITY MEASURED BY AC-SUSCEPTIBILITY OF A YBCO-FILM}

\author{J. K\"otzler, G. Nakielski, M. Baumann, R. Behr, F. Goerke}

\address{Institut f\"ur Angewandte Physik, Jungiusstr. 11,
            D -- 20355 Hamburg}

\author{E.H. Brandt}

\address{Max-Planck-Institut f\"ur Metallforschung, Heisenbergstr. 1,
            D -- 70569 Stuttgart}

\date{Received 18 March 1994}
\maketitle

\widetext

\begin{abstract}
Utilizing a novel and exact inversion scheme, we determine the complex linear
conductivity $\sigma (\omega )$ from the linear magnetic ac-susceptibility
which has been measured from 3\,mHz to 50\,MHz in fields between 0.4\,T and
4\,T applied parallel to the c-axis of a 250\,nm thin disk. The frequency
derivative of the phase $\sigma ''/\sigma '$ and the dynamical scaling of
$\sigma (\omega)$ above and below $T_g(B)$ provide clear evidence for a
continuous phase transition at $T_g$ to a generic superconducting state.
Based on the vortex-glass scaling model, the resulting critical exponents
$\nu $ and $z$ are close to those frequently obtained on films by other means
and associated with an 'isotropic' vortex glass. The field effect on
$\sigma(\omega)$ can be related to the increase of the glass coherence length,
$\xi_g\sim B$.
\end{abstract}


]

%\newpage
\narrowtext

\section{INTRODUCTION}

The widely observed scaling feature of the electrical conductivity in the
presence of a transverse magnetic field $B$ and the universality of the
critical exponents obtained in such scaling analyses on thin
YBa$_2$Cu$_3$O$_7$-films~\cite{1,2,3,4,5,6,7,8,9,10,11} serve as the best
evidences for the existence of a continuous thermodynamic phase transition
to a genuine  superconducting phase occuring at some transition line $T_g(B)$.
Considering  the interplay between the elastic
properties of the vortex lattice and the
disordering potential exerted on the vortices
by density fluctuations of point-like pinning centers on a
microscopic level, the possibility of a glassy vortex state with infinite
barriers for vortex motion, $U(j \rightarrow 0)$ and, hence, zero linear
resistance, $\rho (j \rightarrow 0) = 0$, was argued previously for
3-dimensional lattices (see e.g. Ref.\,12 and references therein). However, a
theoretical proof for the hypothesized second order phase
transition~\cite{13,14} to this vortex glass (VG) which is characterized
by both a singular correlation length and diverging
relaxation time of the order parameter fluctuations,
\begin{equation}  \eqnum{1a} \label{1a}
\xi_g (T,B) = \xi_g(B) | 1-T/T_g |^{-\nu}
\end{equation}
\begin{equation}     \eqnum{1b} \label{1b}
\tau_g (T,B) \simeq \tau_g(B) | 1-T/T_g |^{-\nu z}
\end{equation}
is still lacking. Hence, also the analogy between the transition to the
Meissner-phase at $B = 0$ (relaxational 3d-XY-model) and to the
VG-phase at $B \neq 0$ defining the scaling property of the
conductivity does not have a firm basis yet. Within this model, magnitude
and phase angle are predicted to obey the following forms~\cite{14}
\begin{eqnarray} \eqnum{2a} \label{2a}
| \sigma(\omega) | = \frac{\tau_g}{\xi_g} S_{\pm} (\omega \tau)\,, \\
%\end{equation}
%\begin{equation}
\arctan \left( \frac{\sigma ''}{\sigma '} \right) =
         P_{\pm} (\omega \tau)\,, \eqnum{2b} \label{2b}
\end{eqnarray}
where $S_{\pm} (x)$ and $P_{\pm} (x)$ represent homogeneous
scaling functions above
and below $T_g$, which describe the change from pure ohmic behaviour
$(\sigma ''=0)$ for $T \gg T_g$ to pure screening ($\sigma '=0$)
at $T \ll T_g$. While their
limiting behaviours are known, the detailed shapes of these scaling functions
have not yet been worked out.

This somewhat unsatisfactory theoretical situation contrasts to the ample
evidence for scaling and universality obtained on thin YBCO-films. Most of
these analyses were based on the nonlinear dc-conductivity,
$\sigma (j)$~\cite{1,2,3,4,5,6,7,8},
a recent one on the non-linear Hall resistance~\cite{9}, and
in two studies~\cite{10,11}
the linear dynamic conductivity at $T_g$ has been investigated.
Within the given error
margins, almost all of the published critical exponents can be summarized by
the r.m.s.
averages $\nu$ = 1.7(1) and $z = 5.5(5)$. This represents a nice example for
universality, since most films have been prepared under different conditions
and do
deviate from each other in microscopic composition and
structure, normal resistivity, transition temperature and width.
Moreover, the exponents appear to be insensitive against
variation of  the vortex density  $B/\Phi_0$ and direction
of the field with respect to the c-axis~\cite{7}.

On the other hand, notable exceptions from this universality
have been seen. Dekker et al.~\cite{2} observed a crossover to
two-dimensional-critical behaviour when the film thickness is reduced to
about 16\,{\AA}, i.e. to the thermal correlation length $\xi_T$. I-V data from
microbridges ($80 \times 10 \mu$m$^2$) patterned
from epitaxial YBCO-films have been interpreted~\cite{15} in terms of
thermal activation across Josephson-barriers using the Ambegaokar-
Halperin model~\cite{16}. For YBCO-crystals distinctly different exponents
have been reported (see e.g. Ref.\,17) indicating the possibility of
another universality  class for the transition to true superconductivity
due to the presence of a twin boundary network.

This paper reports a first detailed experimental study of the scaling
behaviour of the linear dynamic conductivity $\sigma (\omega )$ of a thin
YBCO-film in a  wide frequency range covering more than
{\it ten decades} from 3\,mHz to 50\,MHz. This compares to the existing
measurements of
$\sigma(\omega)$ on films between 0.5\,MHz and 500\,MHz~\cite{10,11} which,
moreover, focused only on the critical behaviour of $\sigma(\omega)$
at $T=T_g$. Using experimentally well defined scaling functions for
$\sigma(\omega)$, we will also examine the not yet explained effect of
the magnetic field on the vortex-glass scaling, at least on an empirical
level. Recently, within the mean field approximation\cite{18}, the scaling
behaviour of $\sigma(\omega)$ has been studied on the basis of the
Edwards-Anderson model.

Our experimental method employs a contact-free
access to the dynamic conductivity of a thin film. We measure the linear
dynamic susceptibility $\chi (\omega )$ parallel to both the applied dc-field
and the c-axis, which is normal to the film plane (see section II). Using an
exact inversion scheme~\cite{19} outlined in section III,
we then evaluate from
$\chi (\omega )$ the complex linear conductivity $\sigma (\omega )$ in the film
plane. Section IV presents the scaling analysis of $\sigma (\omega )$ and
determines the relevant results, i.e. the critical exponents, the glass line,
and the field effect on the scaling functions. The results are discussed in
section~V.

\section{EXPERIMENTAL}

The 250 nm thick disk (R=1.5\,mm) under investigation has been cut
from a film, epitaxially grown parallel
to the (001)-axis of a SrTiO$_3$-substrate by pulsed laser deposition
being described in detail elsewhere~\cite{20}. AFM images show a rather
flat surface from which some isolated hills of 1\,$\mu$m average diameter and
up to 0.2\,$\mu$m height grew up. The dc-resistance at 100\,K is about
100\,$\mu \Omega$cm and the transition temperature is $T_c$ = 91.4\,K
with a 10:90$\%$
width of less than 1\,K. From the non-linear ac-susceptibility critical
current densities of $2\cdot 10^6\ $A/cm$^2$ and $10^7\ $A/cm$^2$ were
obtained at 77\,K and 20\,K , respectively.

In order to measure the linear susceptibility, the sample, a thin circular disk
(R = 1.5\,mm) was inserted in well balanced mutual inductances, the excitation
amplitudes of which were always controlled to obtain the linear response.
We use two different detection techniques: between 3\,mHz and 1\,kHz a
commercial SQUID ac-susceptometer (Quantum Design MPMS2) and from 10\,Hz to
50\,MHz conventional lock-in amplifiers (PAR 5301 and PAR 5202).
At fixed external
fields, provided by superconducting magnets in the persistent mode, two
measuring schemes have been used. Most of the data were recorded
by a quasi static variation of the temperature at constant frequencies so that
the same signals were obtained during heating and warming of the sample.
Several measuring curves are reproduced by Fig.\,1. At some interesting points
in the
B-T phase diagram the temperature was stabilized to investigate the frequency
variation towards the low frequency end of our spectrum, where the measuring
time for a magnetic moment of $10^{-7}$ emu to a nominal accuracy of
$\leq 10^{-8}$ emu increased up to 12 hours. In the entire frequency and
temperature range of interest here, the complex output of the mutual
inductances has been carefully calibrated against the known signals of para-
and ferromagnetic samples (Pd, EuO, GdCl$_3 \cdot$\,7\,H$_2$O) of the
appropriate volumes.

\section{AC-CONDUCTIVITY}

  To determine the dynamic susceptibility $\chi(\omega) = (1/2Vh_{ac})
\int {\bf r\times j(r},\omega)\,{\rm d}^3 r$ of a thin disk of thickness
$d$ in a perpendicular ac field $h_{ac} e^{i\omega t}$
one has to solve an integral equation which describes nonlocal
diffusion of the normalized in-plane sheet current density
${\bf J}(r,z) = {\bf j}(r,z) d / h_{ac}$. In the following,
 complete penetration $|\delta | > d$ is assumed, equivalent
to $\omega < 2  /\mu_0 d^2 |\sigma_{ac}|$ with the skin depth
$\delta = (2/\mu_0\omega \sigma)^{1/2}$.
This integral equation follows by inserting the nonlocal relation
between ${\bf J}(r,z)$ and the perpendicular field $h_x(r,z)$
it generates (Amp\`ere's law) into the induction law
${\bf \nabla \times e = -\dot b}$ using the appropriate material laws
${\bf j} = \sigma_{ac}{\bf e}$  and ${\bf b} = \mu_0{\bf h}$.
 For a disk of radius $R$ ($r\le R$) this integral equation reads
\begin{equation} \eqnum{3} \label{3}
  J(r,\omega) = w\Big[  \pi r + \!\int_0^1 \!Q(r,u)\, J(u,\omega)
                \, {\rm d}u \Big]\,,
\end{equation}
where
\begin{equation} \eqnum{3a} \label{3a}
w =  \frac{i\omega R d \mu_0}{2\pi} \sigma_{ac}(\omega)\,.
\end{equation}
In the Ohmic case, $\sigma_{ac} = \sigma$ is
real, thus $w$ is purely imaginary, whereas in the Meissner state or for rigid
pinning, $\sigma_{ac}$ is purely imaginary and thus $w$ is real and
frequency-independent.  The integral kernel is\cite{19}
\begin{equation} \eqnum{3b} \label{3b}
Q(r,u)=-\frac{u}{r} \!\int_0^{\,r/u} \! \Big[ \frac{E(k)}{1-v} +
        \frac{K(k)}{1+v} \Big] v \,{\rm d}v \,.
\end{equation}
Here $E(k)$ and $K(k)$ are complete elliptic integrals and $k^2 = 4v/(1+v)^2$.
Expanding $J(r,\omega)$ in terms of the eigenfunctions
$f_n(r)$ (Fig.\ 2) of the eigenvalue problem
\begin{equation}  \eqnum{4} \label{4}
  f_n(r) = -\Lambda_n \!\int_0^1 \! Q(r,u) \,f_n(u) \,{\rm d}u\,,
\end{equation}
one gets for the planar current density
$J(r,\omega) =\sum_n a_n(\omega) f_n(r)$ with
\begin{equation}  \eqnum{5} \label{5}
  a_n(\omega) =\frac{ \pi\, w\, b_n}{1 +w/\Lambda_n}, \quad
 b_n = \!\int_0^1\! f_n(r) r^2\, {\rm d}r \,,
\end{equation}
yielding the susceptibility
\begin{displaymath}
  \chi(\omega) = \pi\! \sum_n a_n(\omega) b_n \,.
\end{displaymath}
The orthonormality, $ \int_0^1\! f_n(r) f_m(r) r\,{\rm d}r = \delta_{mn}$
of the eigenfunctions was used. The eigenvalues are
$\Lambda_n \approx n - 0.123$ for all $n=1,2,3,\ldots$\,.
Finally, one obtains $\chi(\omega)$ in the form
  \begin{equation}   \eqnum{6} \label{6}
  \chi(\omega) =  \pi\! \int_0^1 \! J(r,\omega)r^2\,{\rm d}r =
  -w\! \sum_n \frac{c_n/\Lambda_n}{\Lambda_n + w} \,,
  \end{equation}
with $c_n = (3\pi^2/8) b_n^2 \Lambda_n^2$. This quite general result expresses
$\chi(\omega)$ as an infinite sum ($n=1\, ...\, \infty$) of terms
which have first order poles at real positions $\Lambda_n$ in the
complex $w$ plane, or at complex positions in the $\omega$ plane.
In the Ohmic case, the poles all lie on the positive imaginary axis
of the $\omega$ plane. If $\sigma_{ac}(\omega)= |\sigma_{ac}| e^{iP}$
is complex, the poles have the phase angles $\pi/2 -P(\omega)$.

Next we show how the amplitudes $c_n$ and eigenvalues $\Lambda_n$
may be calculated and how the infinite sum (\ref{6}) can be
approximated with high precision by a finite sum of $N$ terms
($n=1 \, ... \, N$). After discretizing the continuous variables
 $r$ and $u$ as described in Ref.\,19, the eigenvalue problem
(\ref{4}) is equivalent to the diagonalization of an
$N\times N$ matrix $Q_{ij}$ defined by
\begin{displaymath}
 Q_{i\ne j} = w_j\, Q\,(u_i,u_j) \quad , \quad
    Q_{jj} = w_j\, \ln\frac{0.92363w_j}{2\pi u_j } \,.
\end{displaymath}
Here the $u_i = u(x_i)$ span a nonequidistant grid obtained by
inserting equidistant $x_i = (i-\frac{1}{2})/N$ ($i=1\, ...\, N$)
into an appropriate substitution function, e.g.
$u(x) = (75x -50x^3 +15x^5)/40$ which yields a weight function
$w(x) = u'(x) = (15/8)(1-x^2)^2$ and weights $w_i = w(x_i)/N$.
This grid, and the diagonal terms $Q_{jj}$,
are chosen to maximize the accuracy of the sums, which approximate
integrals like $\int_0^1 \! f(r)\,{\rm d}r \approx \sum\!  f_i w_i$
or $\int_0^1 \! Q(r_i, u) f(u)\,{\rm d}u \approx \sum\!  Q_{ij}f_j$,
see Ref.\ 19. The high accuracy of this grid (error $\sim N^{-3}$)
is achieved by the vanishing of the
weight function $w(x)$ at the edges of the disk, $x=u=r=1$, where
the arbitary function $f(r)$ may be peaked or may have a logarithmically
diverging slope. In fact, the eigenfunctions $f_n(r)$ and the integral
kernel $Q(r,u)$ exhibit such weak singular behaviour at the edge
$r=1$, which is hardly seen on the linear scale of Fig.\ 2.

  The matrix $Q_{ij}$ has exactly $N$ eigenvalues
$\Lambda_n$ and $N$ eigenfunctions $f_n$. Inserting these into
 (\ref{5}) one gets the coefficients $b_n$ and the $N$
amplitudes $c_n$ entering the sum (\ref{6}) for $\chi(\omega)$,
see Table I. The finite sums obtained in this way present the
exact solution of the discretized problem and are very good
approximations to the original problem, the magnetic response
of the disk~\cite{19}.
The sum rules $\sum_n b_n^2 =1/4$, $\sum_n b_n^2\lambda_n = 8/3\pi^2$,
and $\sum_n b_n^2 \lambda_n^{-1} = 4/15$ (see Ref.\ 19)
are satisfied with high precision even for small $N \approx 10$.
However, while for $N\to\infty$ the $b_n^2 \approx 8/\pi^4n^3$
and $c_n\approx 3\pi^2/n$ decrease monotonically with $n$ and the
$\Lambda_n \approx  n-0.123$ are equidistant, finite $N < \infty$
yields nonmonotonic amplitudes $c_n$  and nonequidistant eigenvalues
 $\Lambda_n $ (Table 1) in order to comply with these
sum rules and fit the exact infinite sum (6) for $\chi(\omega)$
in a large range of $\omega$.

In order to invert the measured ac-susceptibilities
$\chi'(\omega) - i \chi''(\omega)$ into ac-conductivities
$\sigma'(\omega)+i\sigma''(\omega)$ by means of the analytical
expression (\ref{6}),
we used a numerical inversion (searching) procedure.
The choices  N = 30 (Table I) and  N = 50 yielded identical
results within the accuracy of our data.

As an example, Fig.\,3 displays the
phase angle and the modulus of $\sigma(\omega)$ evaluated from
$\chi(\omega)$ at 0.4\,T (see Fig.\,1(a)). Even slightly below $T_c$
these data exhibit a frequency dependence which cannot be
explained within the proposed extended models of
thermally activated flux flow (TAFF)~\cite{21}.
The most striking difference occurs in the phase $P$,
which according to TAFF continuously increases from zero
to $\pi/2$ when passing $\omega \approx \tau^{-1}_0 \mbox{exp}(-U/T)$
and then decreases in the same manner to zero above the much
higher frequency $\omega \approx \tau^{-1}_0$ due to
losses originating from the vortex motion within their pinning wells.
This is in stark contrast to the present behaviour (Fig.\ 3)
where, with increasing temperature,  the increase of the phase
$P$ \, ($dP/d\omega > 0$) comes to a halt below $P=\pi/2$ and
is followed by a region with $dP/d\omega < 0$ towards $P=\pi/2$.
Hence our data indicate the onset of collective effects in the fluid
vortex phase near $T_c$, which are not included in the TAFF model.

\section{SCALING  ANALYSES}

When comparing Figs.\,1 and 3, the interesting feature emerges that
$\chi (\omega )$ does not display any direct signal of a thermodynamic phase
transition, whereas the phase $\arctan (\sigma ''/\sigma ')$
of the ac-conductivity
provides strong evidence for such a transition: its frequency derivative
changes sign at some finite value $P$ below $\pi /2$
demanded by the continuity of the scaling function $P_{\pm} (x)$
(Eq.\,(2b)) at $T_g$~\cite{14}. The physical reason behind this
characteristic feature can  be associated with  the
coexistence of infinitely large clusters  $(\xi_g \rightarrow \infty)$
of superconducting glassy and a metallic (liquid)
vortex phase at $T_g$. With increasing temperature, the life time and size of
the
glassy fluctuations decrease so that within larger
observation times $1/\omega $ they
cannot contribute to the screening component $\sigma ''$, whereas towards lower
temperatures the reverse is true: within increasing $1/\omega $ the lossy
contributions $\sigma '$ arising from vortex loop excitations die out at the
expense of the stable glassy phase.

Basing on this signature for the continuous phase transition at $T_g$, the next
step is to check the scaling property predicted for
the phase angle (Eq.\,(2b)) in terms of the variable
$\omega \tau \sim \omega |1-T/T_g|^{-\nu z}$ by a proper adjustment of
the exponent $\nu z$. In fact, Fig.\,4(a) shows that by taking $\nu z$ = 9.35
the data scale very nicely over more than 14 decades in the scaled frequency
$\omega\tau$. Together with the value $z$ = 5.5(5) following from the
phase angle at the transition $P (T_g) = (1-1/z) \pi/2$~\cite{18} also the
exponent $\nu$ is determined, $\nu$ = 1.7(1). The 4\,T-data also shown in
Fig.\,4(a) prove that both exponents are
independent of the magnetic field, at least between 0.4\,T and 4\,T studied
here.  Moreover, these results are
identical with the averages for $\nu$ and $z$, defined by the numerous scaling
analyses of the nonlinear conductivity of thin YBCO-films as outlined in the
introduction. This constitutes an additional independent confirmation of the
universality  feature for YBCO films.
Note that this transition from 'metallic' flux-flow behaviour ($P = 0$)
to true superconductivity ($P = \pi/2$) occurs in rather narrow
temperature intervalls. For 0.4\,T and 4\,T, we find the
$\sigma(\omega)$-data from the intervalls $\pm 0.9$\,K and $\pm 3.0$\,K
around $T_g$ to collapse on the scaling functions. These intervalls
correspond to reduced temperatures
$|1-T/T_g|$ smaller than $10^{-2}$ and $3\cdot 10^{-2}$ as expected for a
critical phenomenon. Interestingly, they are in almost quantitave
agreement with the critical regions quoted by Reed et al.\cite{22} on a
twinned crystal.

Additional support for the scaling property of $\sigma (\omega )$ arises from
the modulus of the conductivity shown in Fig.\,4(b) for the same two vortex
densities.
Here the  natural scaling variable is the dc-conductivity above $T_g$,
$\sigma_g(0) \sim \tau_g/\xi_g \sim |1-T/T_g|^{-\nu(z-1)}$.
Due to the validity of the Kramers-Kronig relations for $\sigma (\omega )$, no
additional parameter is required to determine the scaling according to
Eq.\,(2a).
Using essentially the same glass temperatures and critical exponents,
excellent
scaling of all data from the indicated temperature intervals above and below
$T_g$ is found. Obviously, they collapse on two branches defining
the scaling functions $S_{\pm} (x)$.
Again, their limiting cases $S_{+}(0)$ and $S_{-}(0)$ represent the pure lossy
$(| \sigma | = \sigma ')$ and pure screening
$(| \sigma | \sim \sigma ''\sim \omega ^{-1})$ responses,
while at $T_g$ one finds
the  algebraic behaviour $| \sigma | \sim \omega ^{-1+1/z}$
predicted for the dynamical
criticality. Principally, these results for $|\sigma|$ can be also considered
as a
successful check of the measuring and new inversion procedures presented
here.

The inset to Fig.\,5(b) shows the phase diagram defined by these scaling
analyses.
The transition temperatures define the glass line separating the normal
conducting vortex fluid from the superconducting vortex glass, which can be
described by the power law:
\begin{equation}\eqnum{7} \label{7}
B_g(T) = B_g(0) \left(1-\frac{T}{T_c}\right)^{\beta},
\end{equation}
where $\beta = 1.5(2)$ and $B_g(0) = 85(5)$T.

Looking at the scaling functions $S_{\pm}$ and $P_{\pm}$ exemplified in
Fig.\,4 for 0.4\,T and 4\,T, one notes that their shapes appear rather
similar to each other. This
strongly suggests to perform an additional scaling procedure in a
phenomenological manner. We simply assign suitable powers of the applied
field $B$ to the scaled frequency and conductivity. In fact, Fig.\,5
demonstrates nicely that for both, phase and modulus of $\sigma(\omega)$, all
experimental scaling functions fall on single curves over 15 decades of the
scaled frequency if we assume that the
lifetime $\tau_g$ of the glassy fluctutions increases with $B^{5.5(1.0)}$
and the dc-conductivity, $\sigma(0) \sim \tau_g/\xi_g$ (Eq.\,(2a)), grows
proportional to $B^{4.5(0.5)}$. Interestingly, both results can be related
to the same physical origin, namely to the field dependence of the
vortex-glass correlation length. We obtain the field dependence of the
principal scaling variable in the following form
\begin{equation}\eqnum{8} \label{8}
\xi_g(T,B) = \xi_g(B_0)
             \left( \frac{(B/B_0)^{\gamma}}{|T/T_g -1|} \right)^{\nu},
\end{equation}
where $\gamma$=0.59(8) and $B_0$ denotes an arbitrary normalization field.
This finding implies that the shape of the scaling functions does not
depend explicitly on the magnetic field, but that the field effect
on the scaling can be fully accounted for by the $B$ dependence of the
glass correlation length $\xi_g(T,B)$. Moreover, if we follow Ref. 17 to
express the field dependence of $\xi_g$ in terms of the distance between
$T_g$ and $T_{c2}$, i.e. $|T_{c2}-T_g| \sim T_g B^{\beta'}$, with
$\beta' = 1.6(2)$ for the present glass line, we find within the given error
margins, $\xi_g^{1/\nu}(T,B)\sim (T_{c2}-T_g)/|T-T_g|$. This is fully
consistent with the previous result\cite{17} obtained on a single crystal.

\section{DISCUSSION}

The main information obtained from scaling analyses are the critical
exponents, which are determined by the rather general features of the
phase transition, i.e. the spatial dimensionality,
the symmetry of the order-parameter and the type of disorder.
The present values for the exponents of the
correlation length and the relaxation time,
$\nu=1.7(1)$ and $z=5.5(5)$, respectively, agree nicely
with most of the existing values determined for YBCO-films
\cite{1,2,3,4,5,6,7,8,9} from non-linear I-V characteristics.
On the experimental side, this provides a strong support for the
ac-susceptibility technique and the
evaluation procedure applied here for the first time on a superconducting
thin film. For bulk YBCO-crystals
such an exact transformation of the complex susceptibility to the dynamic
conductivity using conventional
schemes has been performed recently \cite{17}, while another recent scaling
 analysis was performed for the
modulus of $\chi(\omega)$ resting on the assumption
$|\chi(\omega)|\simeq a/|\lambda_{ac}(\omega)|$
\cite{23}. Of course, this assumption is only valid for $|\lambda_{ac}|$
being large compared to the sample dimension
and should be limited to small susceptibilities, i.e. to just
the onset of screening.

The present values of the exponents $\nu$ and $z$ are significantly larger
than the mean-field values\cite{18} for the vortex-glass (VG)
$\nu=0.5$ and $z=4$, indicating the presence of strong
critical fluctuations. For the case of a 3-dimensional isotropic VG
one expects from analogy with spin-glasses \cite{24} larger numbers,
e.g. $z=2(2-\eta)$ with $\eta<0$ \cite{18}, however,
exact values, are not known yet. Monte-Carlo simulations of the
gauge-glass model revealed $\nu=1.3(4)$ and $z=4.7(7)$ \cite{25},
but the applicability to the vortex state in real disordered materials
like YBCO-films has not been proven to date.

On the other hand, the results $\nu = z = 3.1(3)$ from scaling
analyses of $\sigma(\omega)$ for ${\vec B}\parallel {\vec c}$
of twinned YBCO-crystals \cite{17,26} indicate the existence of a
distinctly different universality class for the transition to a superconducting
vortex state. It has been suggested to associate this class with
correlated pinning along the boundaries of twin colonies
($\parallel {\vec c}$), where single twin planes terminate \cite{27}. In this
case of anisotropic pinning, one has to distinguish between two correlation
lengths $\xi_{\parallel}$ and
$\xi_{\perp}$ of glassy fluctuations parallel and perpendicular to the
preferred c-axis. Then the measured
exponents can be explained by the assumption
$\xi_{\parallel}\sim\xi_{\perp}^{2}$.
These anisotropic fluctuations are considered by the so-called
Bose-glass model for a vortex-state with columnar pins\cite{27}, for which
by analogy with the dirty boson problem exists now good evidence for a
phase-transition to genuine superconductivity at a finite temperature
$T_{bg}(B)$\cite{28}. Similar evidence is still lacking for the isotropic
VG-model,
 where the vortices are pinned by
fluctuations of pointlike defects of extension $r_0 \ll \xi_T$.
Interestingly for the YBCO-crystal,
the isotropic values of the exponents $\nu$ and $z$ have been
recovered for ${\vec B}\perp{\vec c}$ \,\cite{17},
i.e. for fields perpendicular to the columnar pinning centers,
which means that these line defects are ineffective in this geometry.
In contrast, when passing from
${\vec B}\perp{\vec c}$ to ${\vec B}\parallel{\vec c}$
for thin YBCO-films, no change of the isotropic exponents
occurs \cite{7}, which reveals an absence of columnar pinning. For our film
such a conclusion for isotropic pinning is supported by the fact that
twin colonies could not be detected down to a length-scale of 0.8 $\mu$m.

Another universal property of VG seems to be the effect of
the magnetic field on the transition. The field dependence of the glass-line
could frequently be explained by the power law, Eq.\,(7), with exponents
similar to the one obtained here, $\beta=1.5(2)$. Usually,
$\beta=2\nu_0$ is related to the exponent of the thermal correlation length
of the homogeneous state\cite{14}, $\xi_T\sim(1-T/T_c)^{-\nu_0}$.
On the other hand, a detailed understanding of the non-universal amplitude
$B_g(0)$, i.e. its dependence on the pinning mechanism\cite{14}
has not yet been achieved.

Much less evidence exists for the field dependence of the dynamic scaling
functions $S_{\pm}$ and $P_{\pm}$ themselves. Our analysis between
0.4T and 4T, shown in Fig.\,5, suggests that their shape is not influenced
by the vortex density, but that the entire effect on the dynamic conductivity
near $T_g$ can be accounted for by that of the VG correlation length.
Our result, Eq.\,(8), implies, that the amplitudes in (Eq.\,(1)),
$\xi_g(B)$ and $\tau_g(B)$, increase approximately linearly with
$B$ (due to $\gamma\nu\approx 1.0(2)$) and with $B^z$, respectively.
Because this finding  agrees with that of our previous study
on an YBCO single crystal for both orientations of the applied
field\cite{17}, it appears to be another rather general property
of the VG-transition. The most obvious consequence of $\xi_g \sim B$ is that
below $T_g$ the squared screening length
$\lambda_s^2$ defined at low frequencies increases linearly with $B$:
\begin{equation}\eqnum{9}\label{9}
\sigma(\omega\to 0,T<T_g,B)=\frac{i}{\mu_0\omega\lambda_s^2}\,,
\end{equation}
\begin{equation}\eqnum{9a}\label{9a}
\lambda_s^2(T,B)=\frac{\lambda_s^2(B_0)}{(1-T/T_g)^{\nu}}
                 \frac{B}{B_0}\,,
\end{equation}
with $\lambda_s(1\mbox{T})=0.7\mu$m.

Let us first note that this behaviour agrees with that of the elastic
pinning length, $\lambda_C^2=B \Phi_0/C_p$,
introduced by Campbell \cite{29} for a vortex lattice exposed to a
pinning energy density $C_p$. In this classical model, the screening
length $\lambda_s$ results from the elastic response of the vortex
lattice and diverges near $T_{c2}(B)$, where the lattice melts. According
to the present results, this melting occurs at $T_g(B)$ and the elastic
response is determined by a field-independent pinning energy $C_p$.

When we associate the screening behaviour, Eq.\,(9) with the VG-fluctuations,
as suggested by the scaling property (Fig.\,5), then the screening
depth $\lambda_s$ is related to the VG correlation length by the form
\begin{equation}\eqnum{10}\label{10}
\xi_g(T,B)\approx\frac{\lambda_s^2(T,B)}{\Lambda_T},
\end{equation}
where $\Lambda_T=\Phi_0/4\pi\mu_0k_BT$ represents the thermal
length\cite{14}. Inserting the experimental $\lambda_s$ we obtain
for the amplitude defined by Eq.\,(1a),
$\xi_g(B)\approx 2.5\mbox{nm}\times B(\mbox{T})$. For the present fields,
this value is in the order of the thermal coherence length at zero
temperature $\xi_T(0)$ which is a natural lower bound to $\xi_g$ at
$T=0$. This implies that, in the critical region,
 $0.002\leq|1-T/T_g|\leq 0.03$, where scaling is
observed here, $\xi_g$ varies from about $1\mu$m to $100\mu$m.
This range is plausible, since it is located
between the vortex-lattice constant, e.g. $a_0\approx 50$nm at
1T and the radius of our disk, $R=1.5$mm, which both set natural bounds
for the range of the fluctuations.

Fisher et al.\cite{14} argued that $\xi_g(B)$ may be associated with the
collective pinning length introduced by Larkin and Ovchinikov\cite{30},
which results from the competition of the elastic and the pinning energies,
$l_p\approx l_i(c_{el}/c_p)^2$, where $l_i$ is the mean distance
between the pinning centers. Taking $\xi_g\sim l_p$ for granted, our result
implies $c_{el}/c_p\sim B^{1/2}$ which is not among the predictions for
$l_p$ in the various pinning regimes\cite{12}. Since the theories of
collective pinning do not considder critical fluctuations but rather apply
to $T \ll T_g$ this discrepancy should not be taken as too serious.

One interesting consequence of our finding $\xi_g \sim B$ is that due
to $\tau_g\sim\xi_g^z\sim B^{5.5}$ the relaxation time of the fluctuations is
growing dramatically with field. In particular, this leads to an increase of
the dc-limit of the fluctuation conductivity,
$\sigma(0)\sim\xi_g^{z-1}\sim B^{4.5}$ as evidenced by Fig.\,5(b).
This is in sharp contrast to the conventional flux-flow behaviour,
$\sigma_{\rm ff}\sim B^{-1}$, and indicates that even slightly below $T_c$ the
dc-limit of $\sigma(\omega)$ in the present YBCO-film is dominated by
cooperative effects. Most likely, this is also true for other films with
glass lines $T_g(B)$ sufficiently close to $T_c$.

Let us finally recall the physical difference between the vortex
dynamics according to the TAFF\cite{31} and related\cite{21} models and
the highly collective VG-model. In the former, single vortices or
bundles become increasingly depinned at elevated temperatures, and this
is described by a single activation rate $\tau^{-1}(T)$. At fixed
frequency, more and more vortices depin within a time $\omega^{-1}$ and
generate electric fields, i.e. ac-losses; this implies that the phase
angle $P=\arctan(\sigma''/\sigma')$, decreases gradually from
$\pi/2$ to zero with
increasing $\omega\tau$. In the VG picture, on the other hand, a flow
of vortex bundles of extension $\xi_g$ occurs only above  the  phase
transition $T_g(B)$.  At $T_g$ bundles of infinite linear extension
appear, which coexist with other, infinite clusters of the vortex-liquid.
Due to this coexistence of both glassy and liquid clusters with
extremely long relaxation rates,
 $\tau_g \sim \xi_g^z$, the phase angle $P$ assumes a nontrivial
 value $(1-1/z)\pi/2$, which lies below the superconducting limit
$P=\pi/2$. This limit is reached for lower temperatures, where the
size $\xi_g(T<T_g)$ of the vortex-liquid clusters, i.e. of the
``metallic'' regions shrinks. Since their lifetime is then strongly
reduced, these clusters do not cause losses at low frequencies
 $\omega < \tau^{-1}$. It is this process which gives rise to the
change of the slope of the phase
$dP/d\omega$ at $T_g$ and is not contained in the TAFF
model\cite{31}. In the extended TAFF-model\cite{21}, a negative slope
of $P(\omega)$ appears at frequencies, where the (viscous) damping of
the pinned vortices is reached. In contrast to the VG-transition, the
crossover from the low-frequency TAFF-regime  ($dP/d\omega > 0$) to the
viscous damping one ($dP/d\omega <0$) is of purely {\it dynamic} origin:
at finite temperatures, there is no diverging time scale, and as long as
the activation energy is large enough, the slope of the phase changes
its sign always at $P = \pi/2$, and the high-frequency limit is
$P=0$ as compared to $(1-z^{-1})\pi/2$ in the VG-model. This dynamic
crossover to the London-screening, where $\lambda_s=\lambda_L$, will be
a matter of future research in the VG-state.

\section{CONCLUSIONS}

In conclusion, we have presented an extensive study of the
linear ac-susceptibility measured perpendicular to
a thin circular YBCO disk, which qualitatively shows
the transition from full penetration to full screening of the ac-field with
increasing frequency and decreasing temperature and field.
In contrast to previous investigations, which discuss
 just the shift of the maximum of $\chi''$ in terms of the TAFF or other
 conventional pinning models,
we have evaluated here the full information available from the data
for $\chi(\omega,T,B)=\chi'- i \chi''$.
This evaluation was achieved by a novel exact inversion routine,
which allows to determine the linear dynamic conductivity
of the bulk material, $\sigma(\omega)$. The frequency dependence of the phase,
$\sigma''/\sigma'$, provides clear evidence
for the existence of a continuous phase transition to a truly superconducting
state at a well defined temperature $T_g(B)$.
Using a step-by-step evaluation of the dynamical scaling proposed for
this transition it was possible to extract the
critical exponents $\nu$ and $z$, which agree with those obtained on a
large number of YBCO-films\cite{1,2,3,4,5,6,7,8,9,10,11}
utilizing the nonlinear dc-conductivity. Since these films differ in
their microscopic structure as well as in their transition temperatures,
the present agreement obtained with a new experimental method, strongly
supports the universality which has been conjectured
for a transition to a vortex-glass \cite{14}.

The dynamical scaling functions for the linear conductivity determined here
for the first time in some detail, perhaps may serve
as a challenge to the theory to work out a realistic model of the
vortex-glass state and its critical behaviour. As a hint for
such an ansatz one may consider the observed linear field
dependence of the coherence length $\xi_g(T,B)$ associated with this
transition. Other interesting questions are the crossovers from the
critical regime to the London-screening on the low-temperature side and to
the flux-flow limit, at $T \gg T_g$.

Acknowledgement: The authors thank M. Kaufmann, M. L\"ohndorf, and
                 U. Merkt (Hamburg) for discussions.

\newpage

\begin{table}
 \caption{Positions of the poles $\Lambda_n$ and amplitudes $c_n$
 entering the transverse susceptibility (6) of a circular disk.
 With these numbers inserted, the finite sum (6) approximates
 the complex function $\chi$ of the complex argument
 $w$ with high precision up to $|w| \approx 4000$.
 These numbers are for $N=30$.}
\begin{tabular}{cccccc}
    $n$ & $\Lambda_n$  & $c_n$ &$n$&  $\Lambda_n$ & $c_n$ \\[2 pt]
    1&  0.8768551 & 0.6355414 & 16 & 14.355486 & 0.0411725 \\
    2&  1.8751352 & 0.2334135 & 17 & 16.420935 & 0.0468693 \\
    3&  2.8725246 & 0.1380313 & 18 & 19.151031 & 0.0530710 \\
    4&  3.8651025 & 0.0965691 & 19 & 22.823289 & 0.0600117 \\
    5&  4.8480195 & 0.0734911 & 20 & 27.880159 & 0.0679791 \\
    6&  5.8147181 & 0.0585680 & 21 & 35.055186 & 0.0773592 \\
    7&  6.7563510 & 0.0479884 & 22 & 45.625953 & 0.0887028 \\
    8&  7.6611613 & 0.0397497 & 23 & 61.958086 & 0.1028391 \\
    9&  8.5126922 & 0.0328696 & 24 & 88.782297 & 0.1210893 \\
    10& 9.2858252 & 0.0264066 & 25 & 136.54610 & 0.1457075 \\
    11& 9.9596476 & 0.0218315 & 26 & 231.60502 & 0.1808932 \\
    12& 10.652940 & 0.0246335 & 27 & 454.24227 & 0.2354904 \\
    13& 11.565606 & 0.0303418 & 28 & 1131.9155 & 0.3319354 \\
    14& 12.125054 & 0.0000002 & 29 & 4536.2877 & 0.5494628 \\
    15& 12.775746 & 0.0357576 & 30 & 81104.991 & 1.5494668 \\
\end{tabular}
\end{table}

\begin{figure}
\caption{Temperature variation of the dispersion and absorption of the
linear magnetic susceptibility measured at fixed frequencies in fields applied
parallel to the c-axis of an YBCO-film :  (a)  B = 0.4\,T   and  (b)  B =
4\,T.}
\end{figure}

\begin{figure}
\caption{Eigenfunctions $f_n(r)$ of the eigenvalue problem, Eq.\,(4),
determining the normalized current density $J(r,\omega)$ of a circular disk.}
\end{figure}

\begin{figure}
\caption{Frequency dependence  (a) of the phase angle $P$ and
(b) of the modulus of the dynamic conductivity $\sigma (\omega)$ evaluated
from the dynamic susceptibility $\chi (\omega)$ measured at 0.4\,T. Full lines
are drawn as guides to the eye. Note that the limit of thermally activated flux
flow, $P = 0$ and  $\sigma (\omega) = \mbox{const.}$, is reached only very
close
to $T_c$.}
\end{figure}

\begin{figure}
\caption{Dynamical scaling of the linear conductivity.  (a) Phase angle and
(b) modulus of $\sigma $, both showing the crossover from pure ohmic to full
screening behaviour at two different magnetic fields.}
\end{figure}

\begin{figure}
\caption{Field-effect on the dynamical scaling functions for (a) the phase
angle and (b) the modulus of $\sigma (\omega)$. The collapse of data is
achieved by introducing appropiate powers of B in the scaling variables.
The inset shows the glass-line (Eq.\,(7)), relative to the mean-field
transition
$B_{c2}$.}
\end{figure}


\begin{references}
\bibitem{1}  R.H. Koch, V. Foglietti, W.J. Gallagher, G. Koren,
             A. Gupta, and M.P.A. Fisher, Phys. Rev. Lett.
             {\bf 63}, 1511 (1989).
\bibitem{2}  C. Dekker, P.J.M. W\"oltgens, R.H. Koch, B.W. Hussey,
             and A. Gupta, Phys. Rev. Lett. {\bf 69}, 2717 (1992).
\bibitem{3}  C. Dekker, W. Eidelloth, and R.H. Koch, Phys. Rev. Lett.
             {\bf 68}, 3347 (1992).
\bibitem{4}  Y. Ando, H. Kubota, and S. Tanaka, Phys. Rev. Lett. {\bf 69},
             2851 (1992).
\bibitem{5}  J. Deak, M. Mc\,Elfresh, R. Muenchausen, S. Foltyn,
             and R. Dye, Phys. Rev. {\bf B 48}, 1337 (1993).
\bibitem{6}  D.G. Xenikos, J.-T. Kim, and T.R. Lemberger, Phys. Rev.
             {\bf B 48}, 7742 (1993).
\bibitem{7}  P.J.M. W\"oltgens, C. Dekker, J. Sw\"uste, and H.W. de
             Wijn, Phys. Rev. {\bf B 48}, 16826 (1993).
\bibitem{8}  J.M. Roberts, B. Brown, B.A. Hermann, and J. Tate, Phys. Rev.
             {\bf B 49}, (1994) preprint.
\bibitem{9}  P.J.M. W\"oltgens, C. Dekker, and H.W. de Wijn, Phys. Rev. Lett.
             {\bf 71}, 3858 (1993).
\bibitem{10} H.K. Olsson, R.H. Koch, W. Eidelloth, and R.P. Robertazzi,
             Phys. Rev. Lett. {\bf 66}, 2661 (1991).
\bibitem{11} Hui Wu, N.P. Ong, and Y.Q. Li, Phys. Rev. Lett.
             {\bf 71}, 2642 (1993).
\bibitem{12} G. Blatter, M.V. Feigel'man, V.B. Geshkenbein,
             A.I. Larkin, and V.M. Vinokur,
             Rev. Mod. Phys. to be published.
\bibitem{13} M.P.A. Fisher, Phys. Rev. Lett. {\bf 62}, 1415 (1989).
\bibitem{14} D.S. Fisher, M.P.A. Fisher, and D.A. Huse, Phys. Rev.
             {\bf B 43}, 130 (1991).
\bibitem{15} N.-C. Yeh, D.S. Reed, W. Jiang, U. Kriplani,
             F. Holtzberg, A. Gupta, B.D. Hunt, R.P. Vasquez,
             M.C. Foote and L. Bajuk, Phys. Rev. {\bf B 45}, 5654 (1992).
\bibitem{16} V. Ambegaokar and B.I. Halperin, Phys. Rev. Lett. {\bf22},
             1364 (1969).
\bibitem{17} J. K\"otzler, M. Kaufmann, G. Nakielski, and R. Behr,
             Phys. Rev. Lett. {\bf 72}, 2081 (1994).
\bibitem{18} A.T. Dorsey, M. Huang, and M.P.A. Fisher, Phys. Rev.
             {\bf B 45}, 523 (1992).
\bibitem{19} E.H. Brandt, Phys. Rev. Lett. {\bf 71}, 2821 (1993);
             Phys. Rev. {\bf B} {\bf 49}, 9024 (1994);
             Phys. Rev. {\bf B} (submitted); and unpublished.
\bibitem{20} M. Schilling, F. Goerke, and U. Merkt, Thin Solid
             Films {\bf 235}, 202 (1993).
\bibitem{21} M.W. Coffey and J.R. Clem, Phys. Rev. Lett. {\bf 67}, 386 (1991)
             and Phys. Rev. {\bf B 45}, 9872 (1992);
             E.H. Brandt Phys. Rev. Lett. {\bf 67}, 2219 (1991);
             Physica C {\bf 195}, 1 (1992).
\bibitem{22} D.S. Reed, N.-C. Yeh, W. Jiang, U. Kriplani, and
             F. Holtzberg, Phys. Rev. {\bf B 47}, 6150 (1993).
\bibitem{23} D.S. Reed, N.-C. Yeh, W. Jiang, U. Kriplani, D.A. Beam,
             and F. Holtzberg, Phys. Rev. {\bf B 49}, 4384 (1994).
\bibitem{24} A.T. Ogielski, Phys. Rev. {\bf B 32}, 5384 (1985).
\bibitem{25} J.D. Reger, T.A. Tokuyasu, A.P. Young, and M.P.A. Fisher,
             Phys. Rev. {\bf B 44}, 7147 (1991).
\bibitem{26} P.L. Gammel, L.F. Schneemeyer, and D.J. Bishop, Phys. Rev. Lett.
             {\bf 66}, 953 (1991).
\bibitem{27} D.R. Nelson and V.M. Vinokur, Phys. Rev. Lett. {\bf 68}, 2398
             (1992) and Phys. Rev. {\bf B 48}, 13060 (1993).
\bibitem{28} M.P.A. Fisher, P.B. Weichmann, G. Grinstein, and D.S. Fisher,
             Phys. Rev. {\bf B 40}, 546 (1989); M. Wallin and S.M. Girvin,
             Phys. Rev. {\bf B 47}, 14642 (1993)
\bibitem{29} A.M. Campbell, J. Phys. C {\bf 2}, 1492 (1969); ibid., {\bf 4},
             3186 (1971).
\bibitem{30} A.I. Larkin and Yu.N. Ovchinnikov, J. Low Temp. Phys. {\bf 34},
             409 (1979).
\bibitem{31} P.H. Kes, J. Aarts, J. van den Berg, C.J. van der Beek,
             and J.A. Mydosh, Supercond. Sci. Technol. {\bf 1}, 242 (1989);
             E. H. Brandt, Z. Phys. B {\bf 80}, 167 (1990);
             Physica C {\bf 185-189}, 270 (1991).
\end{references}
\end{document}